# Flicker Noise in Bilayer Graphene Transistors

Q. Shao, G. Liu, D. Teweldebrhan, A. A. Balandin, S. Roumyantesv, M. Shur and D. Yan

*Abstract*—We present the results of the experimental investigation of the low-frequency noise in bilayer graphene transistors. The back-gated devices were fabricated using the electron beam lithography and evaporation. The charge neutrality point for the transistors was around +10 V. The noise spectra at frequencies *f*>10-100 Hz were of the 1/*f* type with the spectral density on the order of $S_I \sim 10^{-23}$-$10^{-22}$ A$^2$/Hz at the frequency of 1 kHz. The deviation from the *1/f* spectrum at *f*<10-100 Hz suggests that the noise is of the carrier-number fluctuation origin due to the carrier trapping by defects. The obtained results are important for graphene electronic applications.

## I. Introduction

Unique properties of graphene [1-6], such as its extraordinary high room temperature (RT) carrier mobility, up to ~27000 cm$^2$V$^{-1}$s$^{-1}$ [1-3], over 200,000 cm$^2$V$^{-1}$s$^{-1}$ cryogenic mobility [4] and extremely high thermal conductivity in the range of ~3080-5300 W/mK [5-6] make this material appealing for electronic and interconnect applications. For comparison, the RT thermal conductivity of diamond is ~1000 – 2300 W/mK depending on its quality. Many envisioned applications of graphene require low levels of flicker noise, which dominates the noise spectrum at frequencies f below 100 kHz. The flicker noise spectral density is proportional to 1/f$^\gamma$, where γ is a constant close to 1. The unavoidable up-conversion of flicker noise leads to limitations for many electronic devices. It is important to investigate the noise spectrum, its origin and sources in graphene transistors in order to find methods for noise reduction and characterize graphene materials properties.

Gated graphene devices present a unique test bed for investigating the 1/*f* noise origin. Its extremely high electron mobility (limited at RT by the defects rather than phonon scattering) makes graphene particularly well suited for distinguishing between two distinct noise mechanisms: the carrier density fluctuation and mobility fluctuation [7]. In this letter, we report on the fabrication of bilayer graphene (BLG) transistors and analysis of the low-frequency noise in these devices. Based on our results, we suggest a possible noise mechanism and explain its bias dependence. We focus on the BLG transistors for two reasons. Unlike the single-layer graphene (SLG), the BLG has a band-gap [8-9], which is desirable for transistors. Moreover, it was suggested that BLG-based devices may have lower noise levels than SLG devices owing to the specifics of the BLG electron band structure [9].

## II. Device Structure and Fabrication

We prepared graphene layers by mechanical exfoliation from the bulk highly oriented pyrolytic graphite. The chosen method ensured a high quality of the resulting graphene material. The number of graphene layers and their quality were verified using a variety of techniques: optical microscopy, atomic force microscopy (AFM), scanning electron microscopy (SEM) and micro-Raman spectroscopy. The BLG samples were selected via deconvolution of the Raman 2D band and comparison of the intensities of the G peak and 2D band [10-11]. For the device fabrication, we used p-type degenerately doped Si (100) wafers covered with 300 nm thermally grown SiO$_2$. The Si/SiO$_2$ substrates were patterned with the arrays of the square-shaped metal pads with the lengths of 70 μm to 150 μm. The fabrication of pads was performed using lithography with Quintel Q4000 mask aligner followed by 10-nm Cr and 100-nm Au metal deposition by Temescal BJD-1800 electron beam evaporator (EBE). We used Leo1550 electron beam lithography to define the source and drain areas through the contact bars. The 10 nm Cr/ 100 nm Au source and drain contacts were sequentially deposited on graphene by EBE. The bars connected graphene to Cr/Au metal pads. In this design, the degenerately doped Si substrate acted as a back gate.

Figure 1 (a-b) shows SEM and AFM images of the resulting BLG transistors. The dimensions of the source and drain areas defined by the contact bars were ~1.4 μm by 4.0 μm. The channel length was ~9.0 μm and the width of graphene channel was ~ 9.3 μm. The micro-Raman inspection was repeated after the fabrication and electrical measurements to make sure that graphene's crystal lattice was not damaged. Figure 2 presents the spectrum collected from the BLG channel region after all the measurements. The position and intensity of the G peak and

Manuscript received November 04, 2008. The work of Balandin group was supported by DARPA – SRC Focus Center Research Program (FCRP) through its Center on Functional Engineered Nano Architectonics (FENA) and Interconnect Focus Center (IFC), and by AFOSR award A9550-08-1-0100 on the Electron and Phonon Engineered Nano and Heterostructures. The work at RPI was supported by the NRI-funded INDEX Program.

Q. Shao, G. Liu, D. Teweldebrhan and A. A. Balandin are with Nano-Device Laboratory, Department of Electrical Engineering and Materials Science and Engineering Program, Bourns College of Engineering, University of California – Riverside, Riverside, CA 92521 USA (e-mail: balandin@ee.ucr.edu).

S. Roumyantesv and M. Shur are with Center for Integrated Electronics, Department of Electrical, Computer and Systems Engineering, Rensselaer Polytechnic Institute, Troy, NY 12180 USA

D. Yan is with the Center of Nanoscale Sciences and Engineering, University of California – Riverside, Riverside, CA 92521 USA

2D band indicate that the material is indeed BLG, while the absence of D band around 1360 cm$^{-1}$ (indicated by dashed arrow) proves that no substantial damage was introduced during the fabrication.

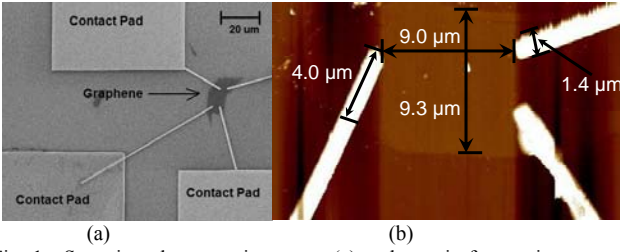

(a)                 (b)

Fig. 1. Scanning electron microscopy (a) and atomic force microscopy (b) images of the back-gated bilayer graphene transistors with the indicated st[...]

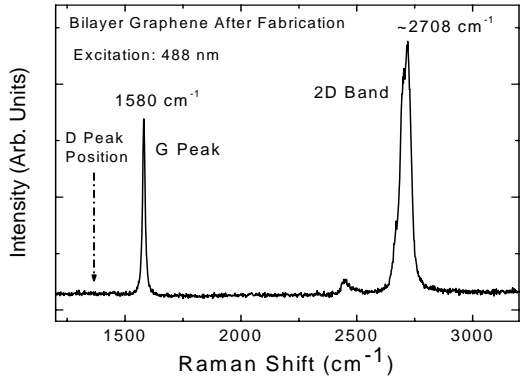

Fig. 2. Raman spectrum collected from the transistor channel region showing features typical for BLG. The absence of the disorder band indicates that no substantial damage was introduced during the fabrication steps

### III. NOISE MEASUREMENTS AND DISCUSSION

The fabricated transistors were kept at ambient conditions for a couple of weeks during the measurements. It was established that the current-voltage (I-V) characteristics were robust and did not change over this period. I-V curves, shown in Figure 3, were measured using the following protocol. First, a positive gate voltage $V_g$ varied from zero to +60 V for a set of the drain-source voltages $V_{ds}$. In this case, the Dirac charge-neutrality point of minimum conductance was observed at $V_g^D$ ~13 V. Next, the negative gate bias swept from $V_g$=0 to -60 V. A small shift appeared at $V_g$=0. The electronic noise was measured in the frequency range from 1 Hz to 10 kHz for a set of biases -20 V < $V_g$ < +53 V and $V_{ds}$ ≤ 60 mV.

Figure 4 presents the noise spectral density measured at $V_{ds}$=20 mV and different gate biases. As one can see, the noise spectral density is on the order of $S_I$ ~ 10$^{-23}$-10$^{-22}$ A$^2$/Hz at $f$=1 kHz for the examined biases, which is a rather low value. The spectral noise density is close to the 1/f noise at frequency $f$>10Hz and small gate voltages. At high gate voltages $V_g$>20V and at $f$<10-100Hz the spectral noise density decreases with the frequency as 1/$f^2$ or even 1/$f^3$. That might be a sign of the generation-recombination (GR) noise with a very small characteristic frequency below 0.1 Hz or of the dc current instability, which manifested itself as slow drift at constant gate and drain voltages on scale of a fraction of a second.

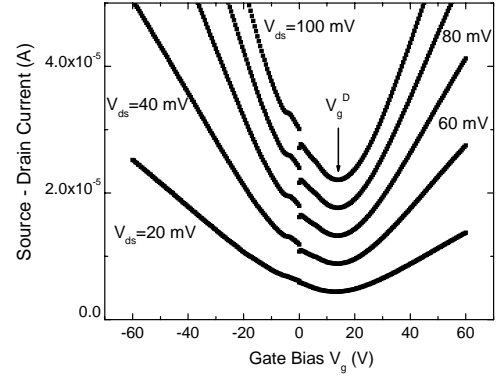

Fig. 3. Current-voltage characteristics of BLG transistor near the charge neutrality point.

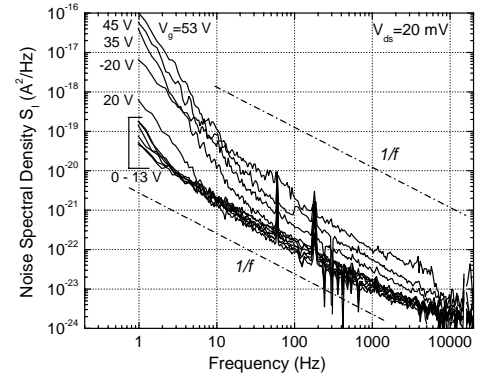

Fig. 4. Noise spectral density as a function of frequency for a range of transistor biases. Note the deviation from the 1/f noise at lower frequencies. 1/f noise spectra are shown for comparison.

In Figure 5, we plot the normalized spectral density $S_I/I^2$ (here I is the source–drain current) as a function of the gate bias for $V_{ds}$=20 mV (triangles) and $V_{ds}$=60 mV (rectangles). The noise level is approximately constant up to $V_g$=20 V, and then increases with increasing gate bias, i.e., with the decreasing electrical resistance of the device channel (see Figure 3). In conventional transistors, the absence of the gate-bias dependence or $S_I/I^2$ growth with increasing $V_g$ would suggest that the noise is dominated by the contributions from the ungated part of the device channel and/or by contacts [12]. If the noise is dominated by the gated channel, its normalized spectral density should decrease with the gate bias [12-13]. Although the conventional formalism can be only applied to the analysis of the back-gated graphene devices with many reservations, our data suggest that the noise contributions from the contacts and/or peripheral regions of the graphene flake are substantial.

The amplitude of the 1/f noise in semiconductors, metals and

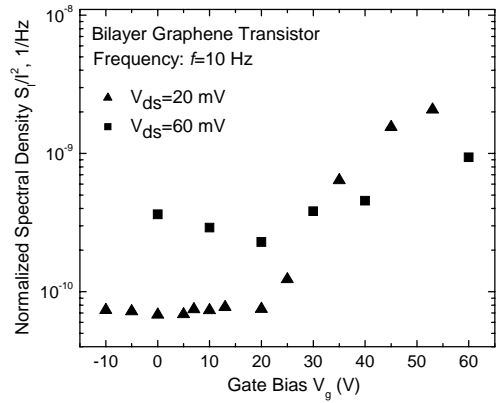

Fig. 5. Normalized noise spectra density as a function of gate bias for two values of the source – drain voltage. The noise level increases with increasing $V_g>20$ V.

semiconductor devices is often characterized by the empirical Hooge parameter

$$\alpha_H = \frac{S_I}{I^2} fN,$$

where $N$ is the total number of carriers in the sample. Since, as discussed above, the contribution of contacts and access regions to measured noise is substantial, we can only get an estimate of the upper bound for Hooge parameter (which can be also affected by a non-uniform electric field distribution for the contact configuration shown in Fig.1). Taking for the total number of carriers $N=L^2/Rq\mu$ (here $L=9$ μm is distance between contacts, $R$ is the resistance, $q$ is the elemental charge and $\mu \approx 3000$ cm$^2$/Vs is the mobility) we obtained $\alpha_H \approx 10^{-4}$ for $f=10$ Hz. Such value of the Hooge parameter is found in good quality semiconductor devices [7].

The deviation of the noise spectral density from the pure $1/f$-type dependence suggests that the low-frequency noise is of the carrier-number fluctuation origin [7]. This type of noise, commonly described by McWhorter model [14], appears as a result of the fluctuation in the number of charge carriers in the channel due to trapping and de-trapping of the carriers by defects. The defects, e.g. lattice imperfections or impurities, can be located inside the graphene bilayer, at the BLG/SiO$_2$ interface, or inside the gate oxide. Since $1/f^{2-3}$ noise was found at very low frequencies, it is dominated by very "slow" traps with the time constants $\tau > 1/(2\pi f_o) \sim 0.3 – 1.1$ sec. This implies that the contributions from the irregular shape peripherals of the graphene flake or from the bulk of the gate oxide, were significant. Based on our analysis, the noise in BLG transistors implemented with the state-of-the-art technology seems to be mostly influenced by the trap distribution and contact quality rather than by the specifics of BLG electronic band structure. The practical implication of this observation is that improvement in the oxide quality, surface passivation and accurate patterning of graphene will be crucial for noise reduction in the few-layer graphene devices.

## IV. CONCLUSIONS

We reported on the fabrication and experimental investigation of the low-frequency flicker noise in the bilayer graphene transistors. The flicker noise in these devices is not of pure $1/f$-type and exhibits $1/f^{2-3}$ behavior at $f <10$ Hz resulting from "slow" traps. The gate bias dependence of the noise density suggests that the contact and graphene edges contributions to the transistor noise were significant. The obtained results are important for the proposed applications of graphene in electronic devices and sensors.